\documentstyle[preprint,revtex,epsf]{aps}
\tightenlines
\begin{document}
\begin{title}Solution to 2+1 Gravity in DreiBein formalism
 \end{title}

\author{W. G. Unruh and Peter Newbury}
\begin{instit}
 CIAR Cosmology Program\\
Dept. of Physics\\
University of B. C.\\
Vancouver, Canada V6T 2A6\\
\end{instit}

 ~

 ~

\begin{abstract}

The reduction of the dreibein formalism of 2+1
General Relativity to the holonomies is explicitly performed.
We also show explicitly how to relate these holonomies to
a geometry classically, and how to generate these holonomies from any
initial data for 2+1 gravity obeying the constraints.
\end{abstract}

\begin{narrowtext}
\section{2+1 Dreibein Formalism}

Witten\cite{1} in a paper a few years ago argued that gravity in the
dreibein formalism was much simpler that the usual geometrical
formalism. He argued that the theory was equivalent to a Chern-Simons
 gauge theory of the 2+1 dimensional Poincare group. His
paper was however at best an ``existence proof", and left many of the
details somewhat vague\cite{2}\cite{3}. The theory was argued to
reduce under the
gauge group to a finite dimensional  theory,
in which the dynamical variables were the Wilson loop
holonomies of a Poincare gauge theory. Although such existence proofs
are of course useful, it can also be instructive to examine the
details of the theory, and to have specific and concrete realisations
to work with. This thought encouraged us to solve the theory
explicitly, so as to see exactly what these holonomies are,
what the conjugate variables are, and what the resultant phase
space looks like. This paper is the report of that investigation.
\cite{Salomonson}

We begin by writing General Relativity in the dreibein formalism,
in which the variables are a set of three vector fields, $e^a{}_\mu$
defined at each spacetime point. $\mu$ is the external
spacetime index, and $a$ is the internal label index.  These fields
 are assumed to form an orthonormal triad, so that
\begin{equation} e^a{}_{\mu} e^{b\mu}=\eta^{ab}\end{equation}
where $\eta^{ab}$ is the flat Minkowski metric with $\eta_{00}=-1$.
 We also then have
that $e^a{}_\mu e^b{}_\nu= g_{\mu\nu}$, the metric on the spacetime.
We will work in the Palatini formalism, in which both the
dreibein and the connection are the fundamental variables. In our
case we will use the internal connection $\omega^a{}_{b\mu}$. The
connection is assumed to be defined so that the internal metric
$\eta_{ab}$ is covariantly constant which gives us that
\begin{equation} \omega_{ab\mu}= - \omega_{ba\mu}\end{equation}
ie, that the connection must be antisymmetric.  (Note that all internal
 indices are raised and lowered by means of the $\eta_{ab}$ metric.)
 In the standard approach,
the bein's are also covariantly constant, so that we want
\begin{equation} \nabla_\mu e^a{}_\nu\equiv e^a{}_{\nu,_\mu}
-\Gamma^\alpha_{\nu\mu}e^a{}_\alpha
+\omega^a{}_{b\mu}e^b{}_\nu=0\end{equation}
Under the assumption that the spacetime connection
$\Gamma^\alpha_{\mu\nu}$ is torsion free, this equation determines
the spacetime connection to be metric compatible, and determines
the internal connection $\omega$ in terms of the derivatives of $e$.
In particular taking the antisymmetric part of the
$\nabla_\mu e^a{}_\nu$, one finds that $\omega^a{}_{b\mu}$ is completely
determined by the ordinary derivatives, $e^a{}_{[\mu,\nu]}$.
(The square brackets as usual denote antisymmetry).
We have
\begin{equation} \omega^a{}_{b[\mu}e^b{}_{\nu]}
=e^a{}_{[\mu,\nu]}.\end{equation}
This equation can be explicitly solved for $\omega^a{}_{b\mu}$.
\begin{equation}
\omega^{ab}{}_\mu ={1\over det(e^f{}_\alpha)}\left( e^d{}_{\nu,\rho}
\epsilon^{\nu\rho\sigma} e_{c\sigma} e_{d\mu}\epsilon^{abc}
- e_{c\mu} \epsilon^{abc} e^d{}_{[\nu,\rho]}
e_{d\sigma}\epsilon^{\nu\rho\sigma}
\right)
\end{equation}
but we shall not need this solution.

Let us define the interior covariant derivative $D_\mu$, such that
if $\rho^a$ is an internal vector on the spacetime, then
\begin{equation} D_\mu\rho^a = \rho^a{}_{,\mu}
+\omega^a{}_{b\mu}\rho^b\end{equation}
On external vectors define it to be simply the ordinary derivative
\begin{equation}
D_\mu v^\nu= v^\nu{}_{,\mu}
\end{equation}
Thus,
\begin{equation} D_\mu e^a{}_\nu= e^a{}_{\nu,\mu}
+ \omega^a{}_{b\mu}e^b{}_{\nu}
\end{equation}

The Einstein action is just given by
\begin{equation} I=\int \sqrt{g}R d^3x\end{equation}
where R is the scalar curvature. Rewriting everything in terms of
the dreibein and the internal connection, one finds that
one can rewrite this action as
\begin{equation} I=\int \epsilon_{abc}\epsilon^{\mu\nu\rho}e^a{}_\mu
\left( \omega^{bc}{}_{\nu,\rho}
+ \omega^a{}_{d\nu}\omega^{cd}{}_\rho \right) d^3x\end{equation}
Here the epsilons are just the antisymmetric symbols with values $\pm 1$ or 0.
The proof is as follows. We define $R^\alpha{}_{\beta\gamma\delta}$ by
\begin{equation}
R^\alpha{}_{\beta\gamma\delta} v^\beta
=2\nabla_{[\gamma}\nabla_{\delta]} v^\alpha
\end{equation}
Multiplying by $e^a{}_\alpha$ we get
\begin{equation}
e^a{}_\alpha R^\alpha{}_{\beta\gamma\delta}e_b{}^\beta e^b{}_\sigma v^\sigma
=2\nabla_{[\gamma}\nabla_{\delta]} e^a_\alpha v^\alpha
=2 D_{[\gamma}D_{\delta]} e^a_\alpha v^\alpha
= R^a{}_{b\gamma\delta}e^b_\sigma v^\sigma
\end{equation}
where $R^a{}_{b\gamma\delta}=2(\omega^a{}_{b[\delta,\gamma]}
+\omega^a{}_{c\gamma}\omega^c{}_{d\delta}).$
Now, defining $e=det(e^a{}_\mu)$, we find that $e=\sqrt{-g}$. Also,
\begin{equation} e_a{}^\alpha=
{1\over 2 e} \epsilon_{abc}\epsilon^{\alpha\beta\gamma} e^b{}_\beta e^c_\gamma
 \end{equation}
or
\begin{equation}
e~e_a{}^{[\alpha}e_b{}^{\beta]}
= \epsilon_{abc}\epsilon^{\alpha\beta\gamma} e^c{}_\gamma
\end{equation}
so
\begin{equation}
\sqrt{-g}R = e~e_a{}^{[\alpha}e_b^{\beta]} R^{ab}_{\alpha\beta}=
\epsilon_{abc}\epsilon^{\alpha\beta\gamma}e^c{}_\gamma
R^{ab}{}_{\alpha\beta}.
\end{equation}

We must check that the equations of motion implied by this action
are identical to those of the original Einstein action. They are. Varying
with respect to the $e^a{}_\mu$ we get the equation of motion
\begin{equation}
 \epsilon_{abc}\epsilon^{\mu\nu\rho}\left(\omega^{bc}{}_{[\nu,\rho]}
+ \omega^{[b}{}_{d[\nu}\omega^{c]d}{}_{\rho]}\right)=0
\end{equation}
which is just the equation
$\sqrt{-g}G^a_\mu=\sqrt{-g} G^\nu_\mu e^a_\nu=0$ written in terms of
the internal connection ($G_{\mu\nu}$ is the Einstein tensor).
 The variation with
respect to $\omega$ leads to
\begin{equation} D_{[\mu} e^a{}_{\nu]}=0\end{equation}
which is just the defining equation for the internal connection to be
metric dreibein compatible.

This action has many appealing features. It is completely polynomial
(in fact is at most quadratic in any one variable). It includes only
the term $e^a{}_\mu$, and not its inverse $e_b{}^\mu$. It is furthermore
linear in the dreibein variables.

As Witten pointed out, the Hamiltonianisation of this action is
trivial because it already is completely in first order in time
derivative form,  Ie, it already is a Hamiltonian action. We take
as our fundamental dynamical variables, $e^a{}_i$ and $\omega^{ab}{}_i$,
the spatial components of the three dreibein, and of the internal
connection. (Indices $ijklmn$ designate the spatial
external spacetime components (ie, 1 and 2), while
 $abcdefg$ and $wxyz$  designate internal
indices. For later use, $pqrst$ will designate the spatial components (1,2)
of internal indices.)  We have
\begin{equation} I
=\int \epsilon_{abc}\left(
e^a{}_1\dot\omega^{bc}{}_2-e^a{}_2\dot\omega^{bc}{}_1
+ e^a{}_0 R^{bc}{}_{12} +\omega^{bc}{}_0( D_1e^a{}_{2}-D_2e^a{}_{1})-
2\partial_{[1}(\omega^{bc}{}_{0} e^a{}_{2]})\right)d^3x
\end{equation}

with the dot representing the derivative with respect to the time,$ x^0$, e.g.,
$\dot\omega^{bc}{}_2= \omega^{bc}{}_{2,0}$.

 We identify
 $e^a{}_1$ as the momentum conjugate to
$\epsilon_{abc}\omega^{bc}{}_2$ and $-e^a{}_2$ as the momentum conjugate to
$\epsilon_{abc}\omega^{bc}{}_1$. There are no momenta conjugate to $e^a_0$ or
 $\omega^{ab}{}_0$, so these act as Lagrange multipliers, and the Hamiltonian
is
 simply the sum of constraints.

The constraints, obtained by varying with respect to the Lagrange multipliers,
 as usual generate  gauge transformations on the variables.
These gauge transformations can in fact be easily integrated, to
give the non infinitesimal gauge transformations. They are of two types.
Firstly, we have the local Lorentz transformations at each point of the
space. This local gauge transformation is
\begin{equation} \tilde e^a{}_i=U^a{}_b e^b{}_i\end{equation}
\begin{equation} \tilde \omega ^{ab}{}_i=U^a{}_{c}U^b{}_{d}\omega^{cd}{}_i
+ U^a{}_{c}U^{cb},_i\end{equation}

Secondly, we have
\begin{equation} \hat e^a{}_\mu = e^a{}_\mu + D_\mu \rho^a\end{equation}
while the connection is left unchanged by this gauge transformation.

This last gauge freedom is intimately connected with the coordinate
transformations of the metric. We have
an infinitesimal gauge transformation
$\hat e^a{}_\mu= e^a{}_\mu+\epsilon D_\mu\rho^a$
so the metric transforms as
$$\hat g_{\mu\nu}=\hat e^a{}_\mu\hat e_{a\nu} =
g_{\mu\nu}+\epsilon(\nabla_\mu e_{a\nu}\rho^a +\nabla_\nu
e_{a\mu}\rho^a)= g_{\mu\nu}+\epsilon 2 \nabla_{(\mu}\beta_{\nu)}$$
which is just the expression for an infinitesimal coordinate
transformation.
However the $\rho^a$ transformation is more general in that one can
choose $\rho$ so as to set $e^a{}_\mu$ equal to zero  in some finite
 region ( and thus the
metric equal to zero) which cannot be accomplished with a coordinate
transformation.

Each of our spaces will be closed, and cannot be covered by a
single open patch without
 overlap. We will assume that the variables are continuous everywhere in the
patch.
We must also specify the relation of the variables on the
 overlap region. We will assume that
there exists some choice of gauge throughout the patch
such that the metric $g_{\mu\nu}$ is well defined  and
unique everywhere. Ie, in the overlap region, the dreibein must be related by
an internal Lorentz transformation.
\begin{equation}
 e_1^a{}_\mu= U^a{}_b e_2^b{}_\mu
\end{equation}
where the 1 and 2 refer to the two overlapping patches.
We furthermore assume that the connection, which is the other set of
fundamental
variables of the theory in the dreibein Palatini formalism, is
implicitly defined by the dreibein, so that if the dreibein's are equal
on two overlapping patches, so are the connections. Although this will turn
out to be true by the equations of motion, we will also assume it true
of the original variables in the sense that if on an overlapping patch
we have
\begin{equation}
 e_1^a{}_\mu= U^a{}_b e_2^b{}_\mu
\end{equation}
then
\begin{equation}
\omega_1^{ab}{}_\mu= U^a{}_c U^b{}_d \omega_2^{cd}{}_\mu +
U^{ac}_{,\mu}U^b{}_{c}
\end{equation}
Ie, across a coordinate patch overlap,  the same gauge transformation relates
both
the dreibein, and the connection.

Note that it would have been simpler to simply demand that the fundamental
variables be continuous everywhere, ie, that the three spatial vector fields
$e^0{}_\mu,~ e^1{}_\mu$, and $e^2{}_\mu$  be chosen continuously. This is
however impossible while still keeping the metric on the
two dimensional slice spacelike.
For any two dimensional surface of genus higher than 1 there exists
no everywhere continuous non-zero vector field defined within the surface.
(``You can't comb the hair on a genus--not--equal--to--unity surface").
Thus, assume that each of the dreibein vector
 fields is continuous, and that at each
point the projection into the surface of the three fields is a basis for the
vectors in the surface. Then at some point the spatial components  of the
second field $e^1{}_i$ must be zero. Thus by assumption,
$e^{0}{}_i$ must be non-zero there, and the ``spatial" metric
 $\gamma_{ij}=-e^0{}_ie^0{}_j + e^2{}_i e^2{}_j$
must have indefinite sign. (ie, the metric is timelike in the surface).
 Thus a dreibein field
on a spatial surface of genus higher than 1 must be discontinuous.

Under the both sets of gauge transformations, the curvature $R^{ab}{}_{\mu\nu}$
transforms simply as an internal  tensor
\begin{equation} \tilde R^{ab}{}_{\mu\nu}
= U^a{}_cU^b{}_d R^{cd}{}_{\mu\nu}\end{equation}
 This
is related to the fact that in addition to being the curvature in the
spacetime, this is also the curvature of the $\omega^{ab}{}_{\mu}$
 regarded as the connection of a gauge theory over the Lorentz group
 of transformations, a fact used by Witten in his paper. Under the second
 set of gauge transformations, since the $\omega$s do not transform,
neither does the  curvature.

Finally under an internal Lorentz transformation, the internal epsilon symbol
\begin{equation}
 \tilde\epsilon_{acb}= U^e{}_aU^f{}_bU^g{}_c\epsilon_{efg}
= det(U)\epsilon_{abc}=\epsilon_{abc}\end{equation}
is left invariant.

We now wish to reduce the above action by applying the constraints,
and performing the appropriate gauge transformations to a fiducial
set of variables. We are going to do so by performing gauge transformations
which reduce certain components of both the connections and the
dreibein to zero on the spatial slice, and we will show that this will
reduce the action to
boundary terms. Although we assume that our spatial slice is a closed
two manifold without boundary,  we will cover the spatial surface by a
coordinate patch, which will have boundaries. The action will reduce to
boundary integrals along the boundaries of this coordinate patch.
 We will finally show how this will
finally reduce the action to one expressed purely in terms of the
holonomies of the connection and the dreibein.

Let us first of all define these coordinate patches. Any closed two topology is
just the connected sum of a series of two dimensional tori. Each two
dimensional
 torus has two
independent cycles. We will choose the edges of our patch to be two
 of these independent cycles. We can furthermore choose the
 edges so that all of the edges
come together at one point. Each cycle is thus
represented in the patch by two edges. The genus n surface
 ( made up of the connected sum of n two dimensional tori) will give a 4n sided
patch. We will use the n=1 and the n=2 as our prime examples but the techniques
we use can be applied for arbitrary n.

Although we will talk about the boundaries of this patch, we will also at times
assume that that the coordinate patch actually overlaps itself at the boundary.

This patch has the property that all curves connecting any two points in the
patch which do not cross the patch boundaries are homotopically equivalent,
 ie can be smoothly deformed into each other. This is certainly not true of
 the n-torus itself, which has an infinite group of homotopically inequivalent
 paths. In fact by construction, each of the patch edges will be a generator
 of this fundamental group of cycles of the genus n surface.

The first constraint is obtained by varying with respect to $e^a{}_0$ and
is
\begin{equation}
\omega^{ab}{}_{[i,j]}+\omega^{a}{}_{c[i}\omega ^{bc }{}_{j]}=0
\end{equation}
This is just the spatial curvature corresponding to this
 internal connection. The gauge transformations generated by these constraints
 are just the
Lorentz transformations. We now  define a Lorentz
transformation at any point on the spatial slice by
\begin{equation} U^a{}_{b}(x)
= \left(P exp(\int_{x_0}^x \omega^\cdot{}_{\cdot i}dx^i)\right)^a{}_{b}
\end{equation}
where the P stands for the path ordering of the  integral,
and the $\omega^\cdot{}_{\cdot i}$ notation means that we are regarding the
internal indices of $\omega$ as matrix indices, and are
appropriately multiplying the matrices in the exponential.
The integral is along some path lying entirely within the
coordinate patch connecting some fiducial point $x_0$ with the
point of interest $x$. Because the spatial curvature of the connection is
zero, $U^a_{~b}(x)$ is well defined and is independent of the path
used to define the integral on the $t=const$ surface.
( Note that because we demanded that
the space be completely covered by the single coordinate patch, and
because that patch has a trivial homotopy, all paths are homotopically
equivalent.) Using this set of Lorentz transformations as a gauge
transformation
on the system sets the spatial components of the transformed connection
 $\tilde \omega^{ab}{}_i$ equal to zero. It does not set the temporal
component to zero.

Because the space is closed,  we must identify points on the
boundary of the patch. Across this boundary $U$ will not necessarily
be continuous. Since an internal Lorentz transformation cannot alter the
metric $g_{\mu\nu}$, the dreibein across the boundary must be related by
some Lorentz transformation. Let us call this transformation $W^a{}_b$.
Ie, if X and Y label the two sides of the identified patch boundary,
\begin{equation}
\tilde e^a_Y{}_i = {W_{YX}}^a{}_{b}\tilde e^b_X{}_i~.
\end{equation}
We have only two vectors ( the two values of $i$) but these are sufficient
 to determine
$W$ up to a time reversal ( assuming that $e^a{}_i$ are spatial vectors).
Demanding that
$W$ lie in the component of the Lorentz group connected to the identity
 therefor uniquely determines  $W$.

The $\omega^a{}_{b\mu}$across the boundary must be related by this
 same Lorentz transformation. But the spatial components of the
$\tilde\omega$ are zero by construction. We therefor have
\begin{equation}
0= 0+{W_{YX}}^{ac}{}_{,i}{W_{YX}}^b{}_c
\end{equation}
from which it immediately follows that $W_{YX}$ must be spatially
constant.

Thus if we use this  $U$ as a gauge transformation, the spatial
components of the
transformed connection will be identically zero
everywhere. The transformed dreibein $\tilde e^a{}_\mu$ is not zero, however.
We now perform the second type of gauge transformation. In particular
we define an internal vector $\rho^a$ by
\begin{equation}
\rho^a(x) = \int_{x_0}^x \tilde e^a{}_\mu dx^\mu.
\end{equation}
Again this is path independent ( for paths within the patch)
because of the constraint obtained by varying the action with respect to
$\omega^{ab}{}_0$.
\begin{equation} 0=\tilde D_{[\mu}\tilde  e^a{}_{\nu]}
= \partial_{[\mu}\tilde e^a{}_{\nu]}\end{equation}
 since $\tilde\omega^{ab}{}_{\mu}=0$.
Using this $\rho^a$ to perform the  gauge transformation of the second type on
$\tilde e^a{}_{\mu}$, we find that the new spatial components of the
dreibein are all equal to zero.

Again, the $\rho^a$ will not be continuous across the boundary, nor are
the $\rho^a$ constant  near the boundary. However, the functions
${\pi_{YX}}^a = \rho^a_Y - {W_{YX}}^a{}_{b}\rho^b_X$ are constants across
 the boundary.
\begin{equation}
0=\tilde e_Y^a{}_i-{W_{YX}}^a{}_b \tilde e_X^b{}_i
 = {\rho_Y}^a_{,i}-{W_{YX}}^a{}_b
{\rho_X}^b{}_{,i}
= ({\rho^a_{Y}}-{W_{YX}}^a{}_b {\rho_X}^b)_{,i}={\pi_{YX}}^a{}_{,i}.
\end{equation}
These $W^a{}_{b}$ and $\pi^a$ are just the holonomies
mentioned by Witten.

How can we now rewrite the action in terms of these holonomies?
The action apparently becomes entirely trivial! Since both the
dreibein and the internal spatial connections are zero, the action
would appear to be zero. This however is because the
naive calculation neglects boundary terms. The action is gauge
invariant under the second type of  gauge
 transformations only up to boundary terms.
Usually one assumes that for a closed universe the boundary terms
are zero. This is not however true in this case.
Clearly, under the Lorentz transformation, the action is left
invariant.

\begin{equation}
\tilde R^{ab}{}_{\mu\nu} = U^a{}_c U^b{}_d R^{cd}{}_{\mu\nu}
\end{equation}
and
\begin{equation}
\tilde I =  \int \tilde \epsilon_{abc}\epsilon^{\mu\nu\rho} \tilde
e^a{}_\mu \tilde R^{bc}{}_{\nu\rho}
= \int\epsilon_{abc}\epsilon^{\mu\nu\rho} e^a{}_\mu R^{bc}{}_{\nu\rho}
\end{equation}

Since the spatial components of the transformed connection are zero, this
leaves just the time component
\begin{eqnarray}
\tilde I&
= \int \epsilon_{abc}\epsilon^{i0j}\tilde
e^a{}_{[i|}\tilde\omega^{bc}{}_{0,|j]}dtd^2x
\nonumber\\
&= \int\tilde\epsilon_{abc}\epsilon^{i0j}
 e^a{}_{[i|}\tilde\omega^{bc}{}_{0,|j]}dt dx^i dx^j
\end{eqnarray}

We now apply the second gauge transformation using our explicitly
constructed $\rho^a$, and use the fact that the $e^a{}_\mu$ obey
the second constraint.
After integrating by parts, using the second constraint, and
writing $e$ in terms of $\rho$, the action reduces to
\begin{equation}
I=- \int_{\partial M}\epsilon_{abc} \tilde\omega^{bc}{}_0 \partial_i\rho^a dx^i
\end{equation}
Now take this boundary integral along the edges of our patch and evaluate the
integrals in pairs. In particular, evaluate the integral
along the  two edges which
are identified on
 the two surface. The identification is direction reversing, so the
expression for the integral along the identified side is the difference between
the
integrals along the two edges of the patch. We use the relation
 between the $\omega^{bc}{}_0$ on the two sides of
the patch  in terms of the $W$.

The $\tilde \omega^{ab}{}_0$ across the boundary are by
 assumption related by the
transformation $W_{YX}$ so that
\begin{equation}
{\tilde\omega_Y}^{ab}{}_0
= {W_{YX}}^a{}_{c}{W_{YX}}^b{}_{d}{\tilde\omega_{X}}{}^{cd}{}_0
+ {\dot W_{YX}}{}^a{}_{c} {W_{YX}}^{bd}.
\end{equation}

Let us look at the contribution to the integral from the two sides labeled X
and Y.
 Since under the
 identification the edges are traversed in opposite directions, we can
 write the integral as
\begin{eqnarray}
 {\int_X \epsilon_{abc} {\rho_X}^a{}_{,i} {\tilde\omega_X}{}^{bc}{}_0}dx^idt &-
{\int_Y \epsilon_{abc} {W_{YX}}^a{}_{d}{\rho_X}^d{}_{,i}
 ({W_{YX}}^b{}_{e}{W_{YX}}^c{}_{f}
{\tilde \omega_{Y}}{}^{ef}{}_0} +{\dot W_{YX}}{}^b{}_{d}{W_{YX}}^{cd}) dx^i dt
\nonumber\\
&={\int_X \epsilon_{abc} ({\rho_X}^a{}_{,i}
\dot{\overline{{W_{YX}}^{-1}}}{}^b{}_{d}{W_{YX}}^{-1}{}^{cd})dx^i}
\\
&={\int_X \partial_i(\epsilon_{abc}{\rho_X}^a
 \dot{\overline{{W_{YX}}^{-1}}}{}^b{}_d {W_{YX}}^{-1}{}^{cd} ) dx^i}
\nonumber
\end{eqnarray}
since the $W$ are constants along the sides.  We get
 for the contribution to the action from these two sides
\begin{equation}
I=...+ \epsilon_{abc}(\rho^a(X_+)-\rho^a(X_-))
\dot{\overline{ W^{-1}_{YX}}}{}^b{}_d W^{-1}_{YX}{}^{cd}
\end{equation}
where the ... denotes the contributions from the other sets of
 identified sides, and $X_+$ and $X_-$ denote the vertices at
 the ends of the side $X$.
Note that the $\dot{\overline{W}} $
notation is used to designate the fact that the dot
 (the time derivative) acts on the whole of the expression under
the bar, i.e., $ \dot{\overline{{UV}}}={d\over dt}(UV) $. The bar
does not denote compex conjugation.

Finally, we can express the values of the $\rho$s at the ends in terms of the
$\pi$s introduced earlier. We will do this explicitly for the
 genus 1 and 2 surfaces and the generalisation to the genus n surface
should should be
 clear.

In order to simplify the notation, we will introduce the following.
Given a set of Lorentz transformations, U and V, let us define the vector
\begin{equation}
\{U\dot V\}_a = \epsilon_{abc}U^b{}_e\dot V^e{}_d U^c{}_f V^{fd}
\end{equation}
Given a vector $\sigma^a$, define the inner product
\begin{equation}
 \{\dot V\}\sigma= \sigma^a\{\dot V\}_a.
\end{equation}
Furthermore, if we Lorentz transform $\sigma$ by $U$, we write
\begin{equation}
U^a{}_b\sigma^b= (U \sigma)^a.
\end{equation}
so we have the inner product between $(U\sigma)^a$ and $\{\dot V\}_a$ as
\begin{equation}
\{\dot V\}_a(U\sigma)^a=\{\dot V\} U\sigma.
\end{equation}
Because the Lorentz transformations have unit determinant, we have
\begin{equation}
\{\dot V\}\sigma=  \{ U\dot V\}U\sigma.
\end{equation}
Because $V^a{}_cV^{bc}=\eta^{ab}$, we also have
\begin{equation}
\{\dot{\overline{UV}}\}\sigma=\{U\dot V\}\sigma +\{\dot U\}\sigma.
\end{equation}
We will use this notation and algebra to greatly
simplify the subsequent calculations.

The genus 1 surface has one rectangular patch that will cover its surface as
 in figure 1.  The edges are labelled A,B,C and D and the vertices
 are numbered 1,2,3 and 4, with the two ends of A having 1
 and 2 as the end vertices, etc. The sides A and C are identified so
 that vertex 1 maps to 4 and vertex 2 maps to 3.  Similarly the sides
 B and D are identified.
We will take the sides A and B as the two
 sides along which we evaluate the action (i.e., they play the role of
 X in the above). The action becomes
\begin{equation} I_1= \{\dot W_{CA}\}(\rho_A(2)-\rho_A(1))
+ \{\dot W_{DB}\}(\rho_B(3)-\rho_B(2))\end{equation}
Since the the value of  $\rho$ at a vertex  depends only on the vertex, not
 the side along which we approach that vertex, we can write
 $\rho_A(1)\equiv \rho_1$, etc, so we have
\begin{equation} I_1=\{\dot W_{CA}\}(\rho_2-\rho_1)+ \{\dot
W_{DB}\}(\rho_3-\rho_2)
\end{equation}

Under identification, all of the vertices meet at a common vertex.
 Figure 2 shows the
relation between the sides and the vertices at the common vertex.
We can now use the relation between the  vectors $\rho$ across an
identified side to
 express all of the $\rho_i$ in terms of $\rho_1$. We have for example
\begin{equation}
\rho_C(3)=W_{CA}\rho_A(2)+\pi_{CA}
\end{equation}
and
\begin{equation}
\rho_D(4)=W_{DB}\rho_B(3)+\pi_{DB}
\end{equation}
But $\rho_A(2)=\rho_B(2)=\rho_2$, so that we can use these two equation to
express $\rho_3$ in terms of $\rho_1$. Similarly across the CA surface we have
\begin{equation}
\rho_4=W_{CA}\rho_1 +\pi_{CA}.
\end{equation}
Finally, again across the DB surface, we have
\begin{equation} \rho_1=W_{DB}\rho_2+\pi_{DB}\end{equation}
which can be used to express $\rho_2$ in terms of $\rho_1$ in an
alternate way.

Substituting the first three of these into the action $I_1$, we get
\begin{eqnarray}
I_1&= \{\dot W^{-1}_{CA}\}(W^{-1}_{CA}W^{-1}_{DB}W_{CA}\rho_1 -\rho_1)
 +\{ \dot W^{-1}_{DB}\}
(W^{-1}_{DB} W_{CA} \rho_1 -W^{-1}_{CA}W^{-1}_{DB}W_{CA}\rho_1)
\nonumber
\\
 &+ \{\dot W^{-1}_{CA}\}(W_{DB}^{-1}W_{CA}^{-1}(\pi_{CA}
-\pi_{DB})-W^{-1}_{CA}\pi_{CA})\\
 & +\{ \dot W_{DB}\}(W_{DB}^{-1}W_{CA}^{-1}(-\pi_{CA}+\pi_{DB})
+W^{-1}_{CA}\pi_{CA}+ W^{-1}_{DB}(\pi_{CA}-\pi_{DB}))
\nonumber
\end{eqnarray}
We can use the algebra of our notation to simplify these. Let us take the
 terms containing $\rho_1$ first.
\begin{eqnarray}
&\{\dot W^{-1}_{CA}\}(W^{-1}_{CA}W^{-1}_{DB}W_{CA}\rho_1 -\rho_1)
 +\{ \dot W^{-1}_{DB}\}(W^{-1}_{DB} W_{CA} \rho_1
-W^{-1}_{CA}W^{-1}_{DB}W_{CA}\rho_1)
\nonumber \\
&=\left( \{W^{-1}_{CA}W_{DB}W_{CA}\dot W^{-1}_{CA}\} -\{\dot W^{-1}_{CA}\}
 +\{W^{-1}_{CA}W_{DB}\dot W^{-1}_{DB}\}
-\{ W^{-1}_{CA}W_{DB}W_{CA} \dot W^{-1}_{DB}\} \right)\rho_1
\nonumber
\\
&= \left(- \{\dot W^{-1}_{CA}\}- \{W^{-1}_{CA} \dot W_{DB}\}
+\{W^{-1}_{CA}  W_{DB} \dot W_{CA}\}
- \{ W^{-1}_{CA}W_{DB}W_{CA} \dot W^{-1}_{DB}\}\right)\rho_1
\nonumber \\
&=\left( \{\dot{\overline{W^{-1}_{CA}  W_{DB}  W_{CA}W_{DB}^{-1}}} \}
\right)\rho_1
\end{eqnarray}
(where we have used
\begin{equation} \{\dot M\}-\{\dot N\}=\{ \dot{\overline{N N^{-1} M}}\} -\{\dot
N\}
=\{N \dot{\overline{N^{-1}M}}\} +\{\dot N\}-\{\dot N\}
= \{N \dot{\overline{N^{-1}M}}\}~~)
\end{equation}
Using the same kind of manipulations on the $\pi$ terms, we finally get
\begin{equation}
I_1=  \{\dot\Omega^{-1}_1\}\rho_1
-\{\dot{\overline{W_{DB}\Omega_1^{-1}}}\}W_{DB}W_{CA}^{-1}\pi_{CA}+
\{\dot{\overline{W_{CA}\Omega^{-1}_1}}\}\pi_{DB}
\label{I1}
\end{equation}
where
\begin{equation}
\Omega_1= W_{CA} W^{-1}_{DB}  W^{-1}_{CA}  W_{DB}.
\end{equation}

We are however not quite finished yet because there are some additional
 constraints on the $W$s and the $\pi$s. In particular, as stated above
there are two
inequivalent ways of expressing $\rho_2$ in terms of the $\pi$s and $\rho_1$.
 These must  be equal to each other for the procedure to be consistent.
 After some simplification, we get
\begin{equation}
 \rho_1
= \Omega_1 \rho_1 +(\Omega_1W^{-1}_{DB}-1)W_{DB}W^{-1}_{CA}\pi_{CA}
 +(1-\Omega_1W^{-1}_{CA})\pi_{DB}\end{equation}
Now $\rho_1 $ is arbitrary, since it depends on the choice of the point $x_0$
from which we begin our integration. Thus the terms dependent and independent
 of $\rho_1$ must be zero separately. We thus get the additional constraints
\begin{equation} \Omega_1=1\end{equation}
and
\begin{equation}
(\Omega_1W^{-1}_{DB}-1)W_{DB}W^{-1}_{CA}\pi_{CA}
+(1-\Omega_1W^{-1}_{CA})\pi_{DB}=0
\end{equation}

Now the variation of the action in eqn. (\ref{I1}) with respect to
 $\rho_1$ just gives us the
 time derivative of the first constraint. That term does not therefor add
 anything new to the equations of motion. We therefor can drop that term.
We can include the constraints into the action by means of Lagrange
 multipliers to finally get
\begin{eqnarray}
I_1&=&-\{\dot{\overline{W_{DB}\Omega_1^{-1}}}\}W_{DB}W_{CA}^{-1}\pi_{CA}
+\{\dot{\overline{W_{CA}\Omega^{-1}_1}}\}\pi_{DB}
\\
&&+ \xi^a{}_b \left(\Omega_1^b{}_a-\delta^b{}_a\right)
 +\left((\Omega_1W^{-1}_{DB}-1)W_{DB}W^{-1}_{CA}\pi_{CA}
+(1-\Omega_1W^{-1}_{CA})\pi_{DB}\right)^a\zeta_a
\nonumber
\end{eqnarray}

However we do have to be a little bit careful. By introducing these
 constraints into the  action we have also introduced the gauge freedoms
 generated by these constraints into the action. We must be certain the the
 original problem actually had these gauge freedoms already, so that we are
 not introducing new gauge freedoms by including the constraints in the action.

There is one gauge freedom which we expect to have, namely that associated
with and overall spatially constant Lorentz
 transformation of everything in the space.
Ie, we expect that for arbitrary Lorentz transformation $L$ such that
$\tilde \pi = L\pi$ and
$\tilde W=LWL^{-1}$, where $L$ is spatially constant,
the action should be left invariant. It is, if we also let
 $\tilde\zeta=L\zeta'$, for an as yet undetermined $\zeta'$,
and $\tilde\xi = L\xi L^{-1}$. Substituting these
 into the action we find that
\begin{eqnarray} \tilde I_1&=~~&
-\{\dot{\overline{LW_{DB}\Omega_1^{-1}L^{-1}}}\}LW_{DB}W_{CA}^{-1}\pi_{CA}
\nonumber
\\
&&+ \{\dot{\overline{LW_{CA}\Omega^{-1}_1L^{-1}}}\}L\pi_{DB}
\\
&&+ (L\xi L^{-1})^a{}_b \left((L\Omega_1 L^{-1})^b{}_a-\delta^b{}_a\right)
\nonumber
 \\
&&+\left(L(\Omega_1W^{-1}_{DB}-1)W_{DB}W^{-1}_{CA}\pi_{CA}
+(1-\Omega_1W^{-1}_{CA})\pi_{DB}\right)^a(L\zeta')_a
\nonumber
\\
&=& I_1-(1-\Omega_1W^{-1}_{DB})W_{DB}W_{CA}^{-1}\pi_{CA}
+ (1-\Omega_1W^{-1}_{CA})\pi_{DB}
+  P_1^a (\zeta'-\zeta-\{\dot L^{-1}\})_a
\nonumber
\\
&=&I_1 + P_1^a(\zeta'-\zeta-\{\dot L^{-1}\})_a
\nonumber
\end{eqnarray}
where
\begin{equation}
P_1=\left((\Omega_1W^{-1}_{DB}-1)W_{DB}W^{-1}_{CA}\pi_{CA}
+(1-\Omega_1W^{-1}_{CA})\pi_{DB}\right)
\end{equation}
which tells us to take
\begin{equation}
\zeta'=\zeta - \{\dot{L^{-1}}\} .
\end{equation}

Which of the constraints generate this gauge freedom? The answer is that
 it is the $P_1$ constraint.  It is instructive to follow through on the
details.

To find what the gauge freedom is that is generated by the $P_1$ constraint,
 we set $\xi$ equal to zero in the action and vary with respect to the
 various dynamical degrees of freedom. Varying with respect to $\pi_{CA}$ we
get
\begin{equation}
  0=-(\{\dot{\overline{W^{-1}_{DB}\Omega_1^{-1}}}\}W_{DB}W^{-1}_{CA})_a
+\zeta_b\left((\Omega_1W^{-1}_{DB}-1)W_{DB}W^{-1}_{CA}\right)^b{}_a
\end{equation}
and with respect to $\pi_{DB}$
\begin{equation}
 0=\{\dot{\overline{ W^{-1}_{DB}\Omega_1^{-1}}}\}_a
+\zeta_b(1-W_{DB}^{-1}\Omega_1^{-1})^b{}_a\end{equation}
where dot is now the derivative with respect to the infinitesimal gauge
 parameter $\tau$.
A solution to these equations (and using the $\Omega_1=1$ constraint) is
\begin{equation}
 \{\dot W_{DB}\}= -\zeta(1-W_{DB})
\end{equation}
\begin{equation}
 \{\dot W_{CA}\}= -\zeta(1-W_{CA})
\end{equation}
Writing $W=L(\tau)W_0L^{-1}(\tau)$, we have
\begin{equation}
\{\dot W\}= \{ W L\dot L^{-1}\} + \{\dot L\}
=\{\dot L\} - \{ W \dot L\}= \{\dot L\} (1-W)
\end{equation}
so taking $\{\dot L\}=\zeta$ solves the equations.

Varying with respect to the $W$s is more difficult because the $W$ are
 Lorentz transformations, not arbitrary matrices. Thus the variations are not
simply
component variations. We define the variation
$\delta W = \Delta W W$ or in components,
 $\delta W^{a}{}_b= \Delta W^{a}{}_{c}W^c{}_b$ where the matrix  $\Delta W$ is
 defined so that
$\Delta W^{ab}=-\Delta W^{ba}$.  This variation preserves
 the Lorentzian nature of the
matrix.
Then
\begin{eqnarray}
&\delta \{\dot W\}_a \pi^a
&= \epsilon_{abc} \left(\dot{\overline{\delta W^b{}_d  }}W^{cd}
+\dot{W}^b{}_d \delta W^{cd}\right)\pi^a
\nonumber\\
&&=\epsilon_{abc}(\dot{\overline{\Delta W}}^{bc}
  + \Delta W^{bd}\dot W_{de}W^{ce}
+ \dot W^{b}{}_{d} \Delta W^{ce}W_{e}{}^d)\pi^a
\\
&&=\dot{\overline{(\epsilon_{abc}\pi^a \Delta W^{bc})}}
-\left( \epsilon_{abc}\dot \pi^a \Delta W^{bc}
+ \epsilon_{abd} \pi^a ( \dot W_{ce}W^{de}
- \dot W^{d}{}_{e} W_c{}^e)\Delta W^{bc}\right)\nonumber
\end{eqnarray}
The first term is a complete  derivative and as usual in the variation
we assume that this is zero ( ie we fix $\pi$ or W at the endpoints of the
 variation). Now,
we have the ansatz that $W=LW_0L^{-1}$ and $\pi=L\pi_0$ so that
$\dot \pi^a=\dot L^a{}_b \pi_0^b= \dot L^a{}_bL^c{}^bL_{cd}\pi_0^d
=\dot L^a{}_bL_c{}^b\pi^c$
But we wanted $\{\dot L\} = \zeta$ or
\begin{equation} \epsilon_{abc}\dot L^{bd}L^c{}_d=\zeta_a\end{equation}
or
\begin{equation}  \dot  L^{bd}L^c{}_d
={{1\over 2}}\epsilon^{abc}\zeta_a\end{equation}
Substituting into the variation of $I_1$ with respect to $W_{CA}$ and
 $W_{DB}$ we find that
\begin{equation} \delta I_1=0\end{equation}
in both cases. Ie, the reputed transformation is exactly generated by the $P_1$
 constraint.

This leaves us with the most interesting constraint, the $\Omega_1=1$
 constraint. The gauge transformations generated by this constraint turn out to
 be simple translations of the value of $\rho$ at all of the vertices.
Ie, the original action is clearly
invariant under the transformation $\rho^a(x)\rightarrow \rho^a(x)+\lambda^a$
where
 $\lambda^a$ is a spatial (but not necessarily temporal) constant since the
action
 depends only on the difference in the values of $\rho$ on the two ends of a
side.
This transformation does however lead to a change in the $\pi$.
 Since the $\rho$s are related across an edge by
\begin{equation}
{\rho_Y}^a={W_{YX}}^a{}_b{\rho_X}^b+{\pi_{YX}}^a
\label{rho}
\end{equation}
a uniform translation of the $\rho $ leads to a change in the $\pi$ of
\begin{eqnarray}
0&=&{\rho'{}_Y}^a-{W_{YX}}^a{}_b{\rho'{}_X}^b-{\pi'{}_{YX}}^a
\nonumber\\
&=&{\rho_Y}^a+\lambda^a -{W_{YX}}^a{}_b{\rho_X}^b-{W_{YX}}^a{}_b\lambda^b
- {\pi'{}_{YX}}^a
\nonumber\\
&=& {\pi_{YX}}^a +(1-W_{YX})^a{}_b \lambda^b-{\pi'{}_{YX}}^a
\end{eqnarray}
or
\begin{equation}
{\pi'{}_{YX}}^a={\pi_{YX}}^a +(1-W_{YX})^a{}_b \lambda^b
\end{equation}
for arbitrary $\lambda^a$.
This gauge transformation is generated by the $\Omega-1$ constraint. (The
proof is straightforward, but we will not present the details here.)

It would now at first seem that we have managed to reduce the problem to a
complete triviality. The set of Lorentz transformations is a three dimensional
 space. Thus the $\Omega$ constraint would seem to put three constraints on
 the dynamical degrees of freedom. The associated gauge transformations
 would then seem to reduce the number of degrees of freedom  by another three.
The $P_1$ constraints would
 seem to be another three constraints, and finally their associated gauge
 transformations would reduce by another three. Thus we would expect to
 reduce the problem by 12 degrees of freedom. But there were only twelve---
 3 for each of the $W$s and three for each of the $\pi$s--- leaving  0
 degrees of freedom. Fortunately this is not the case. The constraints
 are actually degenerate. The $\Omega_1=1$ constraint  can be rewritten as
\begin{equation}
W_{CA}W_{DB}=W_{DB}W_{CA}
\end{equation}
ie, that the $W$ commute. This does not entirely determine one of the $W$s.
 It only says that one must be any arbitrary power of the other.
\begin{equation}
W_{DB}=W_{CA}^\alpha
\end{equation}
for some arbitrary power $\alpha$. Because any Lorentz transformation in
three dimensions always has at least one fixed vector, it always has at
 least one eigenvalue of unity. Because of the $\Omega$ constraint, the
 unity eigenvector of both of the $W$s is the same. Thus the $\pi$ constraints
\begin{equation}
(1-W^{-1}_{CA})\pi_{CA} +(1-W^{-1}_{DB})\pi_{DB}=0
\end{equation}
is actually degenerate in the direction of the fixed vector of the
transformation, leaving only two constraints on the $\pi$s.

In each case, the set of gauge transformations is also only two dimensional,
 leaving a final unconstrained phase space of four dimensions.

Let us now perform a similar reduction in the genus 2 case.
 The procedure, although more
 messy, follows exactly the same procedure as for the
 genus 1 case.

The genus 2 surface has as its patch an octagon as in figure 3. The edges of
	 the octagon we will label A,B,C,D,E,F,G, and H. The vertices we will
	 label with the numbers from 1 to 8. Side A has bounding vertices 1 and 2,
 B has 2 and 3,..., and H has 8 and 1 as bounding vertices. The sides
 A and C are identified so that the vertex 1 maps to 4 and vertex 2 maps
 to 3. B is identified with D with 2 mapping to 5 and 3 to 4, etc. The
 four sides ABCD form one of the connected sum of tori, and EFGH the
 second torus. Again, all eight vertices are actually all mapped to the
 same point in the space. Looking at the connectivity of the sides at
 this common vertex, we get figure 4.
 We will now use this to evaluate the action.

We again make the $U$ transformation obtained from the path
ordered integral of the $\omega^{ab}{}_i$, and the $\rho^a$
 from the integral of $e^a{}_i$. The action reduces to
\begin{eqnarray}
I&
=&\int\epsilon_{abc}\left( (\rho_A{}^a(2)
-\rho_A{}^a(1)){\dot W_{CA}}{}^{b}{}_{d} {W_{CA}}^{cd}\right.
\nonumber
\\
&&~~~~~~~~~+(\rho_B{}^a(3)-\rho_B{}^a(2)){\dot W_{DB}}{}^{b}{}_{d}
{W_{DB}}^{cd}
+(\rho_E{}^a(6)-\rho_E{}^a(5)){\dot W_{GE}}{}^{b}{}_{d} {W_{GE}}^{cd}
\nonumber
\\
&&\left. ~~~~~~~~~~~~~
+(\rho_F{}^a(7)-\rho_F{}^a(6)){\dot W_{HF}}{}^{b}{}_{d} {W_{HF}}^{cd}
\right)dt
\end{eqnarray}
Now, because the $\rho^a$ are continuous on the patch, we have that the
 value of $\rho^a$ at a common end point of two sides is the same.
 So $\rho^a_A(2)=\rho^a_B(2)=\rho_2$, etc. The values of
 $\rho$ across  the boundaries are again given by the relation
 (\ref{rho}) in terms of the $W$s and the $\pi$s, with the equivalent
 set for the vertices 5-8 (see below).

We will group the sides and the vertices into quartets, each of which
will correspond to one of the tori. Thus we group sides ABCD, and take
the vertex 1 as our fiducial vertex. We group EFGH and take vertex 5 as
our fiducial vertex for those sides. The generalisation to higher genus
surfaces is, we hope, clear.

The action is given in the reduced index-free notation by
\begin{equation}
I_2= \{\dot W^{-1}_{CA}\}(\rho_2-\rho_1) +\{\dot W^{-1}_{DB}\}(\rho_3-\rho_2)
    +\{\dot W^{-1}_{GE}\}(\rho_6-\rho_5)+\{\dot W^{-1}_{HF}\}(\rho_7-\rho_6)
\end{equation}
Expressing $\rho$ at the 1234 vertices in terms of $\rho_1$, and at the
5678 vertices in terms of $\rho_5$ we get
\begin{equation}
\rho_4=W_{CA}\rho_1+\pi_{CA}
\end{equation}
\begin{equation}
\rho_3=W^{-1}_{DB}W_{CA}\rho_1 +W^{-1}_{DB}\pi_{CA}- W^{-1}_{DB}\pi_{DB}
\end{equation}
\begin{equation}
\rho_2=W^{-1}_{CA}W^{-1}_{DB}W_{CA}\rho_1 + W^{-1}_{CA}W^{-1}_{DB}\pi_{CA}-
  W^{-1}_{CA}W^{-1}_{DB}\pi_{DB}- W^{-1}_{CA}\pi_{CA}
\end{equation}
\begin{equation}
\rho_8=W_{GE}\rho_5+\pi_{GE}
\end{equation}
\begin{equation}
\rho_7=W^{-1}_{HF}W_{GE}\rho_5 +W^{-1}_{HF}\pi_{GE}- W^{-1}_{HF}\pi_{HF}
\end{equation}
\begin{equation}
\rho_6=W^{-1}_{GE}W^{-1}_{HF}W_{GE}\rho_5 + W^{-1}_{GE}W^{-1}_{HF}\pi_{GE}-
  W^{-1}_{GE}W^{-1}_{HF}\pi_{HF}- W^{-1}_{GE}\pi_{GE}
\end{equation}

We will eventually also need the two relations between $\rho_5$ and $\rho_1$
\begin{equation}
\rho_5=\Omega_1 \rho_1 +(W_{DB} W^{-1}_{CA}W^{-1}_{DB}
-W_{DB} W^{-1}_{CA})\pi_{CA} +(1-W_{DB} W^{-1}_{CA}W^{-1}_{DB})\pi_{DB}
\end{equation}
\begin{equation}
\rho_1=\Omega_2 \rho_5
+(W_{HF} W^{-1}_{GE}W^{-1}_{HF}-W_{HF} W^{-1}_{GE})\pi_{GE}
+(1-W_{HF} W^{-1}_{GE}W^{-1}_{HF})\pi_{HF}
\end{equation}

where we define the $\Omega$s as
\begin{equation} \Omega_1=W_{DB}W^{-1}_{CA} W^{-1}_{DB} W_{CA}\end{equation}
\begin{equation} \Omega_2=W_{HF}W^{-1}_{GE} W^{-1}_{HF} W_{GE}\end{equation}
Note that the relations are entirely symmetric with respect to
 $ABCD\rightarrow EFGH$ and $1234\rightarrow 5678$. The extension
 to higher genus systems is, we hope, obvious.

 Substituting into $I_2$ and simplifying, we get
\begin{eqnarray}
I_2&=& -\{\dot\Omega_1^{-1}\} \rho_1
      -\{\dot{\overline{W_{DB}\Omega^{-1}_1}}\}W_{DB}W^{-1}_{CA}\pi_{CA}
      + \{\dot{\overline{W_{CA} \Omega^{-1}_1}}\}\pi_{DB}
\nonumber\\
&& -\{\dot\Omega_2^{-1}\}\rho_5
      -\{\dot{\overline{W_{HF}\Omega^{-1}_2}}\}W_{HF}W^{-1}_{GE}\pi_{GE}
      + \{\dot{\overline{W_{GE} \Omega_2^{-1}}}\}\pi_{HF}
\end{eqnarray}

Writing $\rho_5$ in terms of $\rho_1$ and simplifying, we get
\begin{eqnarray}
I_2&=& -\{\dot{\overline{\Omega_1^{-1}\Omega_2^{-1}}}\} \rho_1
  -\{\dot{\overline{\Omega_2 W_{DB}\Omega^{-1}_1\Omega_2^{-1}}}\}
\Omega_2 W_{DB}W^{-1}_{CA}\pi_{CA}
      + \{\dot{\overline{\Omega_2 W_{CA} \Omega^{-1}_1\Omega_2^{-1}}}\}
\Omega_2 \pi_{DB}
\nonumber\\
&& -\{\dot{\overline{W_{HF}\Omega^{-1}_2}}\}W_{HF}W^{-1}_{GE}\pi_{GE}
      + \{\dot{\overline{W_{GE} \Omega_2^{-1}}}\}\pi_{HF}
\end{eqnarray}

The constraints are obtained by writing $\rho_1$ in terms of
 $\rho_1$ by going around the vertex.
We get
\begin{equation}
\Omega=\Omega_2\Omega_1=1
\end{equation}
and
\begin{equation}
P_2+\Omega_2 P_1=0
\end{equation}
where
\begin{eqnarray}
P_1&=& (W_{DB} W^{-1}_{CA}W^{-1}_{DB}-W_{DB} W^{-1}_{CA})\pi_{CA}
+(1-W_{DB} W^{-1}_{CA}W^{-1}_{DB})\pi_{DB}
\nonumber\\
&=& ( \Omega_1 W^{-1}_{DB}-1)W_{DB} W^{-1}_{CA}\pi_{CA}
+ (1-\Omega_1 W_{CA}^{-1})\pi_{DB}
\end{eqnarray}
\begin{equation}
P_2=( \Omega_1 W^{-1}_{HF}-1)W_{HF} W^{-1}_{GE}\pi_{GE}
+ (1-\Omega_1 W_{GE}^{-1})\pi_{HF}
\end{equation}

Including the second constraint in the action via a Lagrange Multiplier
again induces the gauge transformation
\begin{equation}
\tilde W= LWL^{-1}~~~~~~~~~;  ~~~\tilde \pi = L\pi
\end{equation}
for all the $W$s and $\pi$s, with a uniform, possibly time dependent L.
The first constraint $\Omega=1$ also again induces the gauge
 transformation corresponding to
a uniform translation of the $\rho$s, which again gives
\begin{equation}
\pi_R = (1-W_R)\lambda
\end{equation}
where the $R$ stands for $CA,DB,GE$,or $HF$, and $\lambda$ depends only on
time.

In the genus 1 case, the constraints plus the gauge freedoms only
reduced the number of degrees of freedom by 8 rather than the expected
 12. In the genus 2 or higher case, they reduce the number by the expected 12.

Let us look at the $\Omega$ constraint first. We can rewrite it as
\begin{equation}
W_{DB}W^{-1}_{CA}W^{-1}_{DB}=\Omega_2^{-1} W^{-1}_{CA}
\end{equation}
Taking the trace of this equation, we get
\begin{equation}
Tr(W^{-1}_{CA})= Tr(W^{-1}_{CA}\Omega_2^{-1})
\end{equation}
Thus given $\Omega_2$, this equation puts one constraint on the possible
Lorentz transformations $W^{-1}_{CA}$. One is left with a two dimensional
set of possible values for $W^{-1}_{CA}$ (see Appendix A).
Only if $\Omega_2=1$ does this
equation not constraint $W^{-1}_{CA}$. For these matrices, $W_{CA}$,
one can find a Lorentz transformation $ W_{DB}$ so
that
\begin{equation}
W_{DB}W^{-1}_{CA}W^{-1}_{DB}= \Omega_2^{-1} W^{-1}_{CA}
\end{equation}

In fact there is a one parameter
family of such transformations, since
 $$LW_{CA}^\alpha W^{-1}_{CA}(LW_{CA}^\alpha)^{-1} = LW^{-1}_{CA}L^{-1}$$
for all powers $\alpha$.  We thus have a two dimensional set of
constraints on $W_{DB}$
 together with the one dimensional constraint on $ W_{CA}$, this
 gives us an effective total of three constraints on $W_{CA}$
and $W_{DB}$ as expected.

 We have argued that for genus greater than 1, the constraint
on the $W$s reduce the number of degrees of freedom by three. Similarly,
the gauge transformations induced by these constraint are also
 three dimensional unless all the $W$s
commute with each other.  We can thus use these three gauge
 transformations to eliminate three of the degrees of freedom
by some sort of gauge fixing.

Similarly, the $\pi$ constraints are also not in general degenerate,
 giving us three constraints and three gauge transformations.
We thus  have 6 constraints and six dimensions of gauge freedom.
 Since, for genus 2, we had 24 degrees of freedom (three each for each of
 $W_{CA},~W_{DB},~W_{GE},~W_{HF}$ and three for each of the $\pi$s
initially), we expect to end up with 12 degrees of freedom finally.
Furthermore in the genus n case we would expect to have 12n-12
degrees of freedom
after all constraints and gauge fixings were carried out. Thus only the
genus 1 case is special, in that the constraints are degenerate.

\section{ Geometry}

What is the relation of these holonomies to the usual 2+1 `metric' approach
to the problem? I.e., in the usual approach, the dynamical
variables are the metric
$\gamma_{ij}$ and the conjugate momentum $\pi^{ij}$. These obey
 three constraints per space point, namely the
 so called ``momentum'' constraint
\begin{equation}
\gamma_{ik}\pi^{kj}_{,j} + \gamma_{ik,j}\pi^{kj}
- {{1\over 2}} \gamma_{jk,i}\pi^{kj}=0
\end{equation}
and the so called Hamiltonian constraint
\begin{equation}
{1\over \sqrt{\gamma}} \pi^{ik}\pi^{jl}(\gamma_{ij}\gamma_{kl}
- \gamma_{ik}\gamma_{jl})
 -\sqrt{\gamma} {}^{(2)}R=0
\end{equation}
where ${}^{(2)}R$ is the two dimensional Ricci scalar
 for the metric $\gamma_{ij}$.

We can find the homologies for any metric and momenta
 which obey these constraints
by the following procedure.

I) Define a dreibein in the following manner. Define $e^0{}_i=0$.
 Choose $e^p{}_i$ ( where the internal indices $pqrst$ will take
values of 1 and 2 only) so that
$\gamma_{ij}= e^p{}_i e_{pj}=e^a{}_ie_{aj}$. The $e^p_i$ are zweibeins
for the two dimensional metric $\gamma_{ij}$. (Raising and lowering of
the 2-internal indices is done via the two dimensional Euclidean
metric $\delta_{pq}$, which is the spatial portion of the three
dimensional Minkowski metric $\eta_{ab}$.)

II) Define the spacetime spatial components of the connection by
\begin{equation}
\omega^{0a}{}_i={1\over \sqrt\gamma}\epsilon^{0ab}\epsilon_{0ij} e_{bk}\pi^{kj}
\end{equation}
\begin{equation}
\omega^{12}{}_i = {1\over \sqrt{\gamma}} e_{ai} e^a{}_{[1,2]}
\end{equation}

III)
Note that because of these definitions that
\begin{equation}
D_{[i} e^a{}_{j]} = 0
\end{equation}
For $a=0$, we have
\begin{equation}
D_{[i} e^0{}_{j]}= e^0{}_{[j,i]} + \omega^0{}_{b[i} e^b{}_{j]}
     = 0 + \epsilon^{0qp}e_{pk}\pi^{kl}\epsilon_{0l[i|}e_{q|j]}
     =-\det(e_{rm}) \epsilon_{0k[j|} \pi^{kl} \epsilon_{0l|i]}
\end{equation}
while for a=p, we have
\begin{eqnarray}
D_{[1}e^p{}_{2]}& = -e^p{}_{[2,1]} - \omega^p{}_{q[i} e^q{}_{j]}
\nonumber\\
&               =-e^p{}_{[2,1]} - \epsilon^{0pq}{1\over \sqrt{\gamma}}
              e_{r[i|} e^r{}_{[1,2]}e_{q|j]}
\nonumber\\
&= e^p{}_{[1,2]} -\epsilon^{0pq}{1\over \sqrt{\gamma}} \epsilon_{0rq}
             \det( e_{sm}) e^r{}_{[1,2]}
\nonumber\\
&=0
\end{eqnarray}
since $\sqrt{\gamma}=det(e_{sm})$.

IV) Finally, we also note that
 \begin{equation} \omega^{ab}{}_{[i,j]} + \omega^a{}_{c[i}\omega^{bc}{}_{j]}=0
\end{equation}
If we choose $a$ or $b$ as 0, this equation follows directly from the
momentum constraint, while if we choose $ab$ as $pq$, this equation
follows from the Hamiltonian constraint.

Thus having defined $\omega^{ab}{}_i$ and $e^a{}_i$ in this way
 we find that they satisfy the requisite dreibein constraints.
We can thus again choose our $U$ and $\rho$ as above, and obtain the
holonomies $W$ and $\pi$. Note that the choice of the hypersurface
enters not at all. The holonomies are entirely defined on any single
spacelike slice on which the dynamic variables obey the constraints.

It is however of interest to note that the holonomies are not defined
on ``superspace", Wheeler's name for the space of spatial geometries.
Neither of the holonomies are functions only of the intrinsic metric.
Rather, the $W$ ( Lorentz) holonomies are functions of both the
extrinsic curvature, through the $\omega^{0p}{}_i$ terms and the
intrinsic geometry through the $\omega^{pq}{}_i$ terms which depend only
on derivatives of the zweibein, which is related to the intrinsic
geometry. The $\pi$ terms are further related to both the intrinsic and
the extrinsic geometries as they are the integral over $U^a{}_b e^b{}_i$
which depend on the intrinsic geometry through the $e^b{}_i$ terms and
the extrinsic geometry through
 $U=P \exp(\int \omega^\cdot{}_{\cdot i}dx^i)$ term.
 If this situation is generic, it would seem that
``superspace" may not be a suitable arena for studying the dynamics of
spacetime.

\section{Quantum Realisation}
In the Dirac quantisation of this system, as stated by Witten,
the simplest system is to
choose the $W$s as our configuration variables, and the $\pi$s
as our momenta. This system is somewhat unusual however in that
 the configuration variables are actually group elements, the
 elements of the Lorentz group. This group has a non-trivial
 structure, in that the space has a non-trivial homotopy---rotations
through $2\pi$ are equivalent to rotations through $0$. It is a
 space with a metric structure induced by the group structure.
 The $\pi$ will induce transformations on this group space.\cite{Nelson}

Let us do this for the genus 1 case, as the simplest case. We
should firstly find a set of canonical coordinates.
Choose $\Pi_{CA}= \pi_{DB}$, $\Pi_{DB}=W_{DB}W_{CA}^{-1}\pi_{DB}$,
$Q_{DB}=W_{DB}\Omega_I^{-1}$, and $Q_{CA}= W_{CA}\Omega_1^{-1}$.
The action is then
\begin{eqnarray}
I_1&=& \{\dot Q_{CA}\}\Pi_{CA} + \{\dot Q_{DB}\}\Pi_{DB}
\nonumber\\
&&+ Tr(\xi(\Omega_1 -1))
 +\zeta \left( (1-Q^{-1})_{CA})\Pi _{CA}+(1-Q^{-1}_{DB})\Pi _{DB}\right)
\end{eqnarray}
The key problem is that expressing $\Omega_1$ in terms of $Q_{CA}$
and $Q_{DB}$ is difficult in general. However, in this particular
case, an equivalent constraint is
\begin{equation} Q_{DB} Q^{-1}_{CA}Q^{-1}_{DB} Q_{CA} =1\end{equation} ,
 which we can use instead. ( Note
that for higher than genus 1, the expression of the $\Omega $ constraint in
terms of the equivalent $Q$ seems to be much more difficult.)
Ignoring the constraints, we find that the $\Pi$ generates an infinitesimal
 Lorentz transformation in the conjugate $Q$. Ie, if we use the same
part of the action with time derivatives, and insert a Hamiltonian
proportional to one of the $\Pi$'s, e.g., $\mu_a\Pi^a$,
we find that the appropriate $Q$ has the equation of motion of
$\{\dot Q\}_a=-\mu_a$, or $\dot Q^a{}_b
={{1\over 2}} \epsilon ^{a}{}_d{}^c \mu_cQ^d{}_b$.
Equivalently, the $Q$ generate a transformation of the $\Pi$s, so
 that a term of the form
$\nu^a{}_b Q^b{}_a$ in the action, creates the motion in the $\Pi$ of the form
$\dot \Pi^a\epsilon_{abc}= (Q\xi)_{[bc]}$. (I.e., the $\Pi$s generate
infinitesimal
Lorentz transformations, while the $Q$s generate translations of the $\Pi$s.)

Because the $\Pi$s generate infinitesimal Lorentz transformations,
we do not expect them to
commute ( since the infinitesimal generators of Lorentz transformations
 do not commute).
 On the other hand, we do expect the infinitesimal
 generators of the translations
( ie the $Q$s) to commute. (By this we mean that $Tr \xi Q$ and
$Tr(\xi' Q)$ will commute as operators, not that $ Q$ and $Q'$ will
commute as matrices). Ie, if we choose the
 $Q$s as our configuration variables, we expect to
be able to find a basis of vectors of our Hilbert
space which are eigenstates of the
Lorentz transformations. Thus we can write our
 wave functions as $\Psi(Q)$, where Q is a Lorentz
transformation. The $\Pi$s will then be the infinitesimal
generators of the Lorentz transformations on this space- ie
\begin{equation} \zeta_a\Pi^a \Psi(Q)
= {d\over d\mu} \Psi(Q^b{}_c+ \mu\zeta_a\epsilon^{ab}{}_d Q^d{}_c)
= \zeta_a\epsilon^{ab}{}_dQ^d{}_c \nabla_{Q^b{}_c}\Psi(Q)
\end{equation}

Note that to set up a representation in the $\Pi$ representation is difficult
because the $\Pi$s will not commute with each other. In fact, from the
representations of the Lorentz group, we know that
\begin{equation} [\Pi^a,\Pi^b] =i \epsilon^{abc}\eta_{cd}\Pi^d\end{equation}
Furthermore, because the Lorentz transformations $Q$ are non-trivial
homotopically
( a spatial rotation through $2\pi$ is the same as no rotation at all),
 the eigenvalues of some of the operators will be discrete. We need to for
example
use $\Pi^0$ and $\Pi^2=\Pi^a\Pi_a$ as the commuting operators to define the
Hilbert space of states.

This suggests that at least in this case, the `metric' basis is in fact a far
more difficult basis in which to work than is the conjugate `connection' basis.

There is further worry  with respect to the quantum system. The
holonomies represent a mapping from the fundamental group of the two
manifold into the Poincare group. However the geometries which
correspond to these Holonomies will have a further discrete invariance.
The geometry after all does not care which of the tori in the direct sum
is labeled the first and which the second, etc. One should expect there
to be an identification of the phase space under relabeling  of the
tori. Furthermore, as Carlip\cite{Carlip-group} points out
in the case of the genus 1
system, the same geometry should result under any homeomorphism of the
fundamental group into itself. This should result in a set of
identifications of the reduced phase space.
Are these identifications even consistent with  the quantum structure
created by the non-commutativity of the $\Pi$s for example? How can one
quantize a system with a compactified identified phase space? These
questions must await further work.

A final comment is on the place of time in this quantum theory\cite{4}.
 This subject will be examined in more detail elsewhere.
 However, as stated, the
reduced action is written purely in term of absolutely conserved quantities.
The completely reduced action then collapses to a trivial action.
 What has happened  tot he time development one expects of geometry,
 etc in this theory? All
of the usual time development is contained in the gauge transformations.
 One could for example choose the time to be just the trace of the
extrinsic curvature, as Moncrief does. One can then write the constants
 of motion in terms of this time, and in terms of some other gauge
dependent dynamical variable (eg, the volume of the space). The Dirac
 wave function $\Psi$ which is a function only of the conserved
 quantities, can now be interpreted as a function of the time
 and the other gauge dependent quantities as well, through the
dependence of the constants of motion on these extra quantities.
 This is however a highly arbitrary procedure. Which gauge dependent
 quantity does one choose as the time, which of the extra quantities
 does one choose in addition to make the constants of motion depend
on? Certainly, without any apriori prejudices, there seems to be  no
 quantity which the theory picks out as in  any sense preferred as the time.

It is possible that if one has a strong prejudice that the
 preferred description of the system is in terms of the spatial
 geometry, that some aspects of the
 extrinsic, or intrinsic geometry may be picked out as special.
 This will
however require further work, because of the tangled and non-linear
 way in which the  geometry is encoded in the holonomies.

\section{Physical Realisation as Geometries}

 Can we explicitly construct a geometry corresponding to a set of
holonomies? This geometry will not be unique as many different
geometries can have the  same holonomies. Furthermore, given the time
development
in the holonomy basis, can we generate a spacetime corresponding to that
time development?
The answer is yes.\cite{Waelbroeck}

Let us say that we have a solution to the equations of motion for the reduced
action- ie we have a set of $W$s and $\pi$s which obey the equations of motion
for some choice of the Lagrange multipliers $\xi$ and $\zeta$. The $\zeta$
terms
result in rigid Lorentz transformations of the $W$s and $\pi$s, while the
$\xi$ terms result in translations of the $\pi$s ( or equivalently of the
$\rho$s).

Now, let $X^a$ be a set of flat coordinates on flat spacetime. Choose
some completely arbitrary set of functions $X_1^a(\tau)$, a path
through the flat spacetime. $\tau$ will be the `time'. Solve the equations
of motion for the holonomies with the given $\xi$ and $\zeta$, and get the set
of $\tau$ dependent $W$s and $\pi$s. At each time, we choose the point
 $X^a_1(\tau)$ as our vertex 1. We chose the other 4n ( where n is the genus)
vertices by using the relations between the $\rho^a$ at the vertices and
applying them to the $X^a$, using the holonomies at time $\tau=0$. Eg, in the
genus 2 case, we define
\begin{equation}
 X^a_4(\tau)= W^a_{CA}{}_b(0) X^b_1(\tau) +\pi^a_{CA}(0)
\end{equation}
\begin{equation}
 X^a_3(\tau)=W_{DB}^a{}_b(0) X^b_4(\tau)+\pi^a_{CA}(0)
\end{equation}
etc., for all 8 vertices.  Having identified the seven vertices, we now
construct
the edges of the spacelike slice, and the surface bounded by these
edges. To make things simpler, we will construct the ``spatial" slices so that
that the spatial geometry
is smooth across the patch boundaries. To do so we want to make sure
that slice is smooth at a vertex, and across each of the edges. We thus
want the edges at the vertex to be coplanar, and we want the surface to
run smoothly across the edges. Let us therefor choose eight coplanar
directions at the vertex 1. We now use the holonomies to move these
eight directions to each of the vertices. We choose these eight
directions to designate the initial directions of each of the eight
sides radiating from their respective vertices. Ie, having found the
initial directions of side A at vertices 1 and 2, we connect vertex 1 and 2
by an arbitrary (spacelike) line with these as the given initial and
final directions.  To find edge  C, we move edge A to C using the $CA$
holonomy.
I.e., ${X_C}^a(\tau)={W^{-1}_{CA}}^a{}_{b}(0) (X^b_A(\tau) -{\pi_{CA}}^b(0))$,
 where ${X_Y}^a(\tau)$
are the coordinates along the edge $Y$ at time $\tau$. Similarly,
 we choose edge $B$ with the given initial and final directions, and find
edge D using the ${DB}$ holonomy, etc.
Now, along edge A,B,E, and F we construct a vector field which is orthogonal to
the edge, and which is coplanar with the edges at the vertices and move this
vector
field to the other edges (C,D,G,H) using the appropriate holonomy.
Having constructed this vector field
along the edges, we now construct the spacelike slice filling in the
figure with this vector field as a tangent vector along each of the
edges. In this way we construct a
hypersurface, which will be smooth everywhere even after the identification
of the appropriate edges, for each value of the time $\tau$. We can now
place coordinates onto each of the identified hypersurfaces, and use
these $\tau ,x^i$ as the coordinates of our spacetime. Locally, these
will form a coordinate system of flat spacetime.
We now wish to construct our vector fields $e^a{}_\mu$. To do so we will define
an appropriate field $\rho^a(\tau,x^i)$. We do so by taking
\begin{equation}
\overline\rho^a(\tau,x^i)= X^a(\tau,x^i)
\end{equation}
  and defining
\begin{equation}
\overline e^a{}_\mu=X^a(\tau,x^i)_{,\mu}
\end{equation}
where $x^\mu= \{\tau,x^i\}$. The metric associated with
$\overline e^a{}_\mu$, ie
\begin{equation}
 g_{\mu\nu}= \overline e^a{}_\mu e_{a\nu}= \eta_{ab}X^a_{,\mu} X^b_{,\nu}
\end{equation}
is clearly just flat spacetime metric in the $\tau,x^i$ coordinates.
Furthermore,
the $\overline e^a{}_{\mu}$ under identifications are related
 by the $W(0)$ transformations
since the $X^a$ and thus the $\overline \rho^a$ are.

Now, the $W(\tau)$ are related to the $W(0)$ by a Lorentz
 transformation
\begin{equation}
W^a{}_b(\tau)= L^a{}_c(\tau) W^c{}_d(0)L_b{}^d(\tau)
\end{equation}
since that is the transformation generated by the
 $\zeta$ term in the Hamiltonian.
Thus define
\begin{equation}
e^a{}_\mu= L^a{}_b \overline e^b{}_\mu.
\end{equation}
These dreibein will be related under identification by the $W(\tau)$.
Define $\rho'{}^a(\tau,x^i)=L^a{}_b(\tau)\overline\rho(\tau, x^i)$
Note that the $e^a{}_i=\rho'{}^a{}_{,i}$, but we no longer have that
$e^a{}_0=\rho'{}^a{}_{,0}$.

Define  $\delta\pi^a(\tau)= \pi(\tau)-L^a{}_b(\tau)\pi^b(0).$
There will exist
some vector $\sigma^a(\tau)$ such that
$\delta\pi_{YX}^a(\tau)= (\delta^a_b-W^a_{YX}b(\tau))\sigma^b(\tau)$
because this is the transformation generated by the $\xi$ term in the
Hamiltonian.
Thus we can finally define
\begin{equation} \rho^a(\tau,x^i)=L^a{}_b(\tau)X^a(\tau,x^i)
+ \sigma^a(\tau)\end{equation}
Note that neither the $L$ transformation, nor the $\sigma$ displacement change
the geometry of the spacetime at all. They are simply internal transformations
which
change the representation of the geometry in terms of $e^a{}_\mu$ and
the representation of the $e^a{}_\mu$ in terms of the $\rho^a$.

As far as the spacetime geometry is concerned, the holonomies are simply
constants of the motion--- ie, one can choose a representation such that
they are simply constants.

One feature of this construction, is that the slices are not necessarily
spacelike.
 Depending on the value on $X^a_1(\tau)$ used as the
initial vertex 1, the various vertices of the patch can be either timelike
or spacelike separated from each other. Furthermore one can arrange the
evolution to take the system from one of the spacelike slices, to a
timelike slice. For large values of $X^a(\tau)$, the W transformations will
dominate in determining the location of the other vertices.  The
location of the other vertices scales like $X^a_1$ for the $W$ transformations,
but
the $\pi$ translations are independent of the value of $X^a_1$. Thus,
if $X^a_1$ is large and timelike separated from $0$, the vertices will all
 approximately lie on the spacelike hyperboloid determined by
$X^aX_a=X^a_1X_{1a}$.
If $X^a_1$ is spacelike separated from 0, they will all approximately lie on a
timelike hyperboloid, and the spacetime will have a closed timelike curves.
There appears to be nothing to prevent one from choosing the $X^a_1(\tau)$ from
going
from a large spacelike to a large timelike value, giving a
transformation from a regular spacelike surface to one with closed timelike
curves.

Let us illustrate this with a genus 1 example. Take
\begin{equation}
 {W_{CA}}^a{}_b=\left(\matrix{ \cosh(\theta)&\sinh(\theta)&0\cr
                     \sinh(\theta)&\cosh(\theta)&0\cr
                      0&0&1\cr} \right)
\end{equation}
\begin{equation} W_{DB}=1\end{equation}
\begin{equation} \pi_{CA}=0\end{equation}
\begin{equation} \pi_{DB}=(0,0,1)\end{equation}
We will take $X^a_1=(\tau,1,0)$, and will take as our internal
coordinates $\phi,z$, with the identification $\phi+\theta=\phi$ and
$z+1=z$. The slices will be given by
\begin{equation}
 X^a(\tau,\phi,z) = ( \cosh(\phi)\tau+\sinh(\phi),
\cosh(\phi)+\sinh(\phi)\tau, z)
\end{equation}
with
\begin{equation}
e^a{}_0= (\cosh(\phi),\sinh(\phi),0)
\end{equation}
\begin{equation}
e^a{}_1=(\cosh(\phi)+\tau\sinh(\phi), \tau\cosh(\phi)+\sinh(\phi),0)
\end{equation}
\begin{equation}
e^a{}_2=(0,0,1)
\end{equation}
with metric defined by
\begin{equation}
d^2s= -d\tau^2 - 2 d\tau d\phi + (\tau^2-1)d^2\phi +d^2z
\end{equation}
with the above identifications. This metric is regular everywhere
 except for $\tau=0$, when the spatial slices overlap. In particular it is
regular
when the surface becomes timelike, and the identifications produce closed
timelike
curves at $\tau=1$. It is furthermore of interest that the geometry obtained by
choosing the curve
\begin{equation} X'^a_1=(\tau',0,0)\end{equation}
and $\phi',z$ defined as before produces the  geometry
with metric
\begin{equation}
d^2s= -d^2\tau' +\tau'^2 d^2\phi' +d^2z.
\end{equation}
The two metrics are in fact equivalent. The transformation
\begin{equation} \tau'^2=\tau^2-1\end{equation}
and
\begin{equation}
 \phi'= \phi + {{1\over 2}}
\ln\left( {\tau+1\over \tau-1}\right)
\end{equation}
takes one into the other, Furthermore, the $\tau=const$ surfaces
 map into each other, while the identification,
$\phi\equiv \phi+\theta$
implies
$\phi'=\phi'+\theta$. This illustrates the fact that these geometries in
 general
contain closed timelike curves.

 We would expect this to be a generic
feature of all of these closed 2+1 geometries. We would expect them all
to contain closed
timelike curves. We can always take $X_1$ to be large and spacelike.
 In that case the $W$s will translate the points along a timelike surface
( the points with the same length $X^aX_a$ as $X_1$.) Since the action of
 the $W$s form
 a closed figure in this two surface, some of the vertex points will
 have to be timelike separated from others, giving us closed timelike
 curves. Since $X_1$ is assumed to be sufficiently large, the $\pi$s
 cannot alter this conclusion.

The only difficulty with the argument is that the spacetime may become
singular on the way from the large timelike $X_1$ to the large spacelike
one. In particular there are axes in the spacetime which are mapped into
themselves ( possibly with a translation) by the holonomies. These axes
will be conical singularities in the spacetime. The question one can ask
 is whether the spacetime, evolved from regular initial data,
meets these singularities "before"  it develops closed timelike
curves. ( I.e., whether or not there is a slicing of the spacetime which
 is non singular, which contains a regular initial data surface, and which
 contains a slicing with closed timelike curves, but is non-singular for
 all slices between the regular slice and the one containing closed
timelike curves.)

The complete generic answer to the above question is a subject for
future work, but
 some simple examples can be given to show that the answer may be yes, that
closed timelike (or null) curves  are a generic feature of the 2+1
 spacetimes even in this developmental sense.
 For the simplest regular genus 2 surface,
 the regular octagon
with the Lorentz holonomies defined below, and with the $\pi$s all zero,
it is difficult to propagate through the null cone without singularity. It
is however possible to show that there is a non-singular almost
everywhere spatial two surface in the spacetime with a closed null curve
lying in the regular two surface. Ie, the spacetime does develop closed
temporal curves "before" it goes singular (if it does). To see this in the case
of the $\pi=0$ spacetimes, consider one of the Lorentz transformations,
call it $V$, which connects one of the vertices to one of the other vertices
 ( say $X_1$ to one of the other vertices). Assume that this is a
Lorentz boost, rather than a rotation ( the appendix shows that not
 all of them can be rotations). The boost $V$ will be around some
 spacelike axis. Choose $X_1^a$ to lie perpendicular to this axis
 and to be null. (This is always possible as there is always a null
vector orthogonal to any spacelike vector.) Then the coordinates of the
other vertex obtained from $X_1$ by $V$ will be a null vector, and will
 lie on the same null radial direction from $0$ as does $X_1$. I.e.,
the difference vector between the two coordinates will also be null. Thus
one has a null line connecting two identified points in the space, and it
is therefor a closed null curve. The other vertices all lie on the null
 cone as well, and are in general all spacelike separated from each other
(any two points on the null cone are either spacelike or
null separated from each other). One will therefor be able to
 fill in the surface between these
vertices with a spacelike or null surface.

The only  holonomies for which we might not expect such closed timelike
curves are
 those for which all of the $W$ are the identity. In that case however,
 the surface is essentially n flat tori, joined only at a single
point, since  $X_{4r+1}$ must be the same as $X_1$ in that case.

 One could also arrange situations in which the vertex 5 and
vertex 1 became identified at some time, with
regular spatial slices before ( or after) this singular time. This might
be taken as a model of bifrucation of the initial genus 2 surface into two
 genus 1 surfaces.  All we would need would be to choose our
holonomy so that for some value of $X_1^a$,
\begin{equation}
X_5=\Omega_1 X_1 +(W_{DB} W^{-1}_{CA}W^{-1}_{DB}-W_{DB} W^{-1}_{CA})\pi_{CA}
+(1-W_{DB} W^{-1}_{CA}W^{-1}_{DB})\pi_{DB}=X_1.
\end{equation}

We have done this in figures 5-7 for the regular octagon---i.e., the
genus 2 surface with regular $W$ homologies so that choosing $X_1$
appropriately will give the other vertices spaced as a regular octagon.
($W_{CA} $ is chosen to be a boost about the $(-1.234,1.123,-1.123)$
 axis with a boost
angle of 2.2568. $W_{DB}$ is the same about an axis rotated ( in the $xy$
plane) by $45^o$, $W_{GE}$ is the same rotated by $180^o$ and $W_{HF}$ is
$W_{CA}$ rotated by $225^o$. We have chosen the $\pi$s appropriately, so
that at $\tau=1$, where $\tau= (X_1^aX_{1a})^{1\over 2}$, $X_5=X_1$.
Before this time, as in figure 6, the parts of the figure overlap, and
we have closed timelike curves. At the time of figure 5, the loop around
the space running from vertex 5 to vertex 1 has zero length. The point
is actually a cone, with a Lorentz boost deficit angle (just $\Omega_1$
or $\Omega_2$ depending on which of the two section you are looking at
the point from.) Ie, the space has degenerated into two tori, joined at
one point, each having a conical singularity there. In figure 7, at an
even larger value of $\tau$, the space has opened into a perfectly well
behaved  spatial 2 surface. The pinching off of figure 5 gives one
the possibility of a topology change from a genus 2 to two genus 1
surfaces, especially in the quantum regime.

These examples suggest that in 2+1 gravity, the issue of closed timelike
 curves will have to be handled, and raises the possibility that similar
issues could also arise in four dimensional gravity. In particular the
dreibein ( or fierbein) formulation seems to be ignorant of ( or seems
able to handle, depending on your point of view) situations which one
would, in the usual geometric context, consider highly pathological.
The seeming ubiquity of closed timelike curves in three dimension illuminates
another facet of the issue of time in quantum gravity. The reduced variables
are labels on spacetimes, and more importantly on spacetimes which in some
 regions behave well, and in others appear to be very badly behaved (from
the point of view of one interested in the time development of the world).
This is an issue which will clearly require more work to understand.

The possibilities are numerous, and we would expect that the 2+1
problem will remain a fruitful laboratory for explorations in
quantum gravity for a while yet. It does however provide us with a laboratory
 in which many proposals for quantum gravity in 3+1 dimensions can be tested,
 precisely because the usual ADM formulation of 2+1 gravity mimics that in
3+1 dimensions. The theory has the grave disadvantage
(from the viewpoint of reality) that the local dynamics
is so trivial ( although this is not apparent in the ADM
 formulation), and the concomitant advantage that it is
exactly solvable. It thus expands our repretory of model
 systems, beyond the usual mini-superspace cosmological
models, in which we can test our ideas for quantum gravity.
 In particular, it allows us to test our notions of the
importance of and the role of time in a quantum theory of gravity.

\acknowledgements
We thank Steve Carlip for helpful discussions and for pointing out some of
 the work in the field we had missed. WGU thanks NSERC and the CIAR for
 partial support while this work was being done.

\section{Appendix A}
We want to show that the equation
\begin{equation}
ABA^{-1}B^{-1}=C
\end{equation}
places three constraints on $A$ and $B$. To do so we will work in the
two dimensional representation of the Lorentz group. In this representation,
 the generators of the  group are given by multiples of the usual Pauli
 matrices, $\Sigma_0=\sigma_3$, $\Sigma_1=i\sigma_1$ and
$\Sigma_2=i\sigma_2$. They obey
\begin{equation}
\Sigma_a\Sigma_b= -\eta_{ab}+i\epsilon_{abc}\Sigma^c
\end{equation}
A generic Lorentz rotation is represented by
$$U=exp(i\theta\hat u^a \Sigma_a) = \cos\theta +i \sin(\theta){\hat
u^a}\Sigma_a$$
where $\hat u^a$ is a unit positively directed timelike vector.
 The angle $\theta$ runs from $-\pi$ to $\pi$, because in the
 two dimensional representation of the Lorentz group, the angle
 is half the usual rotation angle.

A generic boost is given by
$$U=exp(i\hat u^a \Sigma_a) = \cosh|u| + i\sinh(|u|){\hat u^a}\Sigma_a$$
where $|u|= \sqrt{ u^au_a}$, and $u^a$ is a spacelike vector.
 ( Note that $|u|$ is ${1\over 2}$  the usual Lorentz  boost parameter).

To find the action of a Lorentz transformation on a vector $v^a$, we write
$$\tilde v^a\Sigma_a = U(v^a\Sigma_a) U^{-1}.$$
$\tilde v$ will be a Lorentz transformed vector.

We will write the constraint in this representation.
The constraint equation now has the form
\begin{equation}
 ABA^{-1}=CB
\end{equation}
Taking traces, we get
\begin{equation}
Tr(B)=Tr(CB)
\end{equation}
We now have four possibilities, depending on whether
$B$ abd $C$ are boosts or rotations (The null cases do
 not add anything essential to the discussion).

A) Both boosts:

We have the trace condition
\begin{equation}
\cosh(|b|) = \cosh(|c|)\cosh(|b|) +\sinh(|c|)\sinh(|b|) {b_a c^a\over |b||c|}
\end{equation}
or
\begin{equation}
\tanh(|c|/2) =-\tanh(|b|) {\hat b_a \hat c^a}
\end{equation}
where $\hat b_a$ is the unit vector parallel to $b_a$.
 Since $\tanh$ has a
 maximum value of unity for real arguments, we must have
$0<{tanh(|c|/2)\over \hat b_a\hat c^a}<1$ for a solution to exist.
 Thus, we can choose $\hat b^a$ almost arbitrarily, and always find
 a value of $|b|$ that obeys this trace constraint. This ensures
that the magnitude of the vector $b^a$ is the same as the magnitude
of the vector $\tilde b^a$ associates with the matrix $CB$
$$CB= e^{\tilde i b^a\Sigma_a}$$.
Now, given any spacelike vector, there always exists a Lorentz
transformation which takes it into any other spacelike vector with
the same magnitude. Thus there always exists a Lorentz transformation
 which  takes $b^a$ to $\tilde b^a$, and thus a Lorentz matrix A which
 satisfies the constraint. Although the direction of $b^a$ is restricted
 by the inequality, it still leaves a largely arbitrary choice for the
direction of $b^a$.

B) B a rotation, C a boost.

The trace constraint is
\begin{equation}
\cosh(|c|)\cos(\theta_b) +\sinh(|c|)\sin(\theta_b)(\hat c^a\hat b_a)=
cos(\theta_b)
\end{equation}
or
\begin{equation}
\tanh(|c|/2)= -\tan(\theta_b) \hat b_a\hat c^a
\end{equation}
Since the dot product between a spacelike and a timelike vector goes as
 $\sinh \theta$ for some $\theta$, and since $\tan\theta$ can have
 arbitrary  value, we find that this equation always has a solution
 for $\theta_b$ for any $\hat b^a$. However we must check one further
 condition. In order that $ABA^{-1}=CB$, $CB$ must have the same sign
of rotation as does $B$, i.e., $\theta_b$ and $\theta_{\tilde b}$, defined by
$$CB= e^{i\theta_{\tilde b}\tilde b^a\Sigma_a}$$
must have the same sign since a Lorentz transformation cannot change
the sign of the rotation angle. Ie, the term multiplying $\Sigma_0$
in $CB$ must have the same sign as the term multiplying $\Sigma_0$ in $B$.

We have the $\Sigma$ dependent terms of CB
$$ \cosh(|c|) \sin(\theta_b) \hat b^a\Sigma_a
+ \cos(\theta) \sinh(|c|) c^a\Sigma_a
- \sinh(|c|) \sin(\theta_b)\epsilon_{abc}\hat c^a \hat b^b\Sigma^c $$
Without loss of generality we can take $c$ to point in
the $1$ direction. Looking at the $\Sigma_0$ component, we have
$$ cosh(|c|) sin(\theta_b) \hat b^0
+sinh(c^1)sin(\theta_b)\epsilon_{120} \hat b^2
= sin(\theta_b)(b^0 cosh(c^1)+b^2 sinh(c^1))$$
Thus this has the same sign as $sin(\theta_b)b^0$, the
$\Sigma_0$ term in $B$. Thus  there will exist an A transformation
 which will take B to CB. We can thus choose the direction of $b^a$
 arbitrarily, as long as it remains timelike.

C) C rotation, B boost:

\begin{equation}
\tan(\theta_c/2)=\tanh(|b|) \hat b^a \hat c_a
\end{equation}
Since $\tanh(|b|)$ is positive and bounded by 1, we again get a
 constraint not only on the  value of $|b|$ but also on the direction of $b$ of
\begin{equation}
0<{\tan(\theta_c/2)\over \hat b^a \hat c_a} <1
\end{equation}
For any $b^a$ which obeys these two constraints ( one an inequality),
we can always find a Lorentz transformation $A$ to
 rotate $b^a$ into $\tilde b^a$.

D) Both rotations.

\begin{equation}
\tan(\theta_c/2)=\tan(\theta_b) \hat b^a \hat c_a
\end{equation}
Since $\hat b^a \hat c_a<0$, the only constraint is that the sign of
$\theta_b$ must be opposite that of $\theta_c$. For all directions
 $\hat b^a$ one can always find a magnitude $\theta_b$ which solves
 this constraint.

However we must again check the $\Sigma_0$ term.
 We can take $c^0=1,~~c^i=0$ without loss of generality. We obtain for the term
multiplying $\Sigma_0$
\begin{equation}
\cos(\theta_c)\sin(\theta_b) b^0 + \cos(\theta_b)\sin(\theta_c)
 =(\tan(\theta_c)+b^0\tan(\theta_b))\cos(\theta_c)\cos(\theta_b)
\end{equation}
But from the constraint, we have that $\tan(\theta_b)
= -\tan(\theta_c/2)/b^0$. so the $\Sigma_0$ term becomes
\begin{equation}
cos(\theta_b)\cos(\theta_c)( tan(\theta_c)-\tan(\theta_c/2))
\end{equation}
which has the same sign as $\theta_c$, and which must be opposite
the sign of $\theta_b$ to solve the constraint. There is no regular
 Lorentz transformation which can change the sign of the angle.
Thus one can never find a rotation $B$ and some matrix $A$ which
 will solve the constraint with a rotation $C$.

\end{narrowtext}

\newpage
\centerline{\bf Figure Captions}
\centerline{Figure 1}
\centerline{ \it The coordinate patch for a genus 1 surface.}

\vskip .1in
\centerline{Figure 2}
\centerline{\it The location of the vertices and edges at the common
vertex of the genus 1 surface}
\vskip .1in
\centerline{Figure 3}
\centerline{\it The patch for a genus 2 surface with the edges and
 vertices labeled.}
\vskip .1in
\centerline{Figure 4}
\centerline{\it The common vertex with the edges and vertices
labeled for a genus 2 surface.}
\vskip .1in
\centerline{Figure 5}
\begin{quote}
{\it The regular octagon genus 2 surface with additional
non-zero
 $\pi$s to make vertices 1 and 5 coincide. Drawn at time
$\tau=1$. The vertices have been joined with straight lines, which would
make the identified figure have kinks at the joining surface, and at the
common vertex. These could, however, be smoothed out by the techniques
described in the text.}
\end{quote}
\vskip .1in
\centerline{Figure 6}
\begin{quote}
{ \it The same spacetime as in Figure 5, but at time
$\tau=.5$. Note that the sides in this $xy$ projection have overlaps
indicating  closed timelike curves.}
\end{quote}
\vskip .1in
\centerline{Figure 7}
\begin{quote}
{\it The same spacetime as in figure 5 but at a later time of
$\tau=1.5$. Note that now the slice is a regular spatial slice. }
\end{quote}

\end{document}